\documentclass{article}
\usepackage{emulateapj}
\usepackage[american]{babel}

\def\spose#1{\hbox to 0pt{#1\hss}}
\def\lta{\mathrel{\spose{\lower 3pt\hbox{$\mathchar"218$}}
     \raise 2.0pt\hbox{$\mathchar"13C$}}}
\def\gta{\mathrel{\spose{\lower 3pt\hbox{$\mathchar"218$}}
     \raise 2.0pt\hbox{$\mathchar"13E$}}}

\def\clock{\count0=\time \divide\count0 by 60
     \count1=\count0 \multiply\count1 by -60 \advance\count1 by \time
     \number\count0:\ifnum\count1<10{0\number\count1}\else\number\count1\fi}

\submitted{Accepted for publication in the Astronomical Journal}

\lefthead{Yan et al.}
\righthead{Extremely Red Objects from the NICMOS/HST Parallel Imaging Survey}
 
\begin{document}

\title{Extremely Red Objects from the NICMOS/HST Parallel Imaging Survey}

\bigskip

\author {Lin Yan\altaffilmark{1}, Patrick J. McCarthy\altaffilmark{1},}

\author{Ray J. Weymann\altaffilmark{1}, 
Matthew A. Malkan\altaffilmark{6},
Harry I. Teplitz\altaffilmark{4,5},
Lisa J. Storrie-Lombardi\altaffilmark{1,2},
Malcom Smith\altaffilmark{3},
Alan Dressler\altaffilmark{1}}

\altaffiltext{1}{The Observatories of the Carnegie Institution of Washington, \\
                  813 Santa Barbara St., Pasadena, CA 91101; Email: lyan@ociw.edu}

\altaffiltext{2}{SIRTF Science Center, Caltech, Pasadena, CA 91101}

\altaffiltext{3}{Cerro-Tololo Interamerican Observatory, Casilla 601, La Serena,
Chile}

\altaffiltext{4}{NASA Goddard Space Flight Center,
                  Code 681, Greenbelt, MD 20771 }

\altaffiltext{5}{NOAO Research Associate}

\altaffiltext{6}{Astronomy Department, University of California,
                 Los Angeles, CA 90024-1562 }


\begin{abstract}
We present a catalog of extremely red objects discovered using
the NICMOS/HST parallel imaging database
and ground-based optical follow-up observations.
Within an area of 16 square arc-minutes, we detect 15
objects with $\rm R - F160W > 5$ and $\rm F160W < 21.5$.
We have also obtained K-band photometry for a subset of the 15 EROs.
All of the $\rm R - F160W$ selected EROs imaged at K-band have $\rm R - K > 6$.
Our objects have $\rm F110W - F160W$ colors in the range of 1.3 $-$ 2.1, redder than
the cluster ellipticals at $z \sim 0.8$ and nearly 1 magnitude redder
than the average population selected from the F160W images at the same depth.
In addition, among only 22 NICMOS pointings, we detected two groups or clusters
in two fields, each contains 3 or more EROs, suggesting that
extremely red galaxies may be strongly clustered.
At bright magnitudes with $\rm F160W < 19.5$, the ERO surface density
is similar to what has been measured by other surveys. At the limit of our sample,
F160W $= 21.5$, our measured surface density is 0.94$\pm 0.24$ arcmin$^{-2}$. Excluding the
two possible groups/clusters and the one apparently stellar object, reduces the
surface density to 0.38$\pm 0.15$~arcmin$^{-2}$.

\end{abstract}

\keywords{Galaxies --- Elliptical and Starburst galaxies}

\section{Introduction}
Deep near-IR imaging surveys have revealed a population of extremely
red objects (``EROs''; Elston, Rieke \&\ Rieke 1988; McCarthy, Persson
\&\ West 1992; Graham \&\ Dey 1996; Hu \&\ Ridgeway 1994; Soifer et al.
1994; Dey, Spinrad \&\ Dickinson 1995; Thompson et al. 1999).  The
nature of the extremely red population remains unclear. As it is defined
largely by a single color, primarily $\rm R - K$, there is no certainty
that it represents a uniform class of object, and it may contain
contributions from galaxies, cool stars or substellar objects and
active nuclei.  The precise definition of an ERO varies among the
different surveys and depends on the particular bandpasses employed.
Most samples were defined by $\rm R - K \gta 5 - 6$ or $\rm I - K \gta 4
- 5$.  The present work is based on a somewhat different color system,
that defined by the NICMOS F160W bandpass and conventional Kron-Cousins
R magnitudes. The NICMOS F160W band is similar to the Johnson
H band filter and the $\rm H - F160W$ color term is negligable for a
flat spectral energy distribution (in $f_{\nu}$ units)
(M. Rieke, 1999, private communication).
Among the resolved objects there are reasonable
expectations that there should exist stellar systems at redshifts such
that the  K-correction applied to an old or intermediate age population
will produce very red optical to near-IR colors. Alternatively even
fairly modest extinction, when observed in the same redshift range, can
produce very steep spectral energy distributions in the rest-frame
near-UV and there are local examples of such objects among the dusty
starburst population.  The central issue regarding the nature of the
resolved EROs is to understand to what degree these two classes of
objects contribute to the overall population.

The earliest interpretation of the colors of EROs were centered around
the old stellar population hypothesis (e.g. McCarthy, Persson, \& West
1991; Hu \& Ridgeway 1994) and redshifts of 1 - 2 were inferred from their
multi-band photometry. There are clear examples which support
this interpretation, such as weak radio source LBDS 53W091
at $z = 1.55$ (Dunlop et al. 1996) and near-IR selected
object CL0939+4713B at z$=1.58$ (Soifer et al. 1999), and a concentration of EROs
at $z = 1.3$ (Liu et al. 2000).
Graham and Dey (1996) argued that the spectral energy distribution of
HR10 (Hu \& Ridgeway 1994) is
well matched by a dusty star-forming galaxy at $z = 1.5$.  The
detection of a strong sub-mm continuum from HR10 (Cimatti et al. 1998;
Dey et al. 1999) provided conclusive
evidence that some of EROs, if not all, are dust enshrouded starburst galaxies
with inferred star formation rates of $\rm 500-2000 h_{50}^{-2}M_\odot~yr^{-1}$ at
moderate redshifts ($z \sim 1 -2$).  Recent deep near-IR follow-up
observations of the sub-mm sources detected with the SCUBA (Smail et
al. 1998) have suggested that two faint sub-mm sources may also be EROs
with $\rm I - K > 6$ (Smail et al. 1999).
Liu et al. have measured redshifts for
several other EROs and most of these are also in the $0.8 < z < 1.5$
range and some have moderately strong emission lines (Liu et al. 2000).

Two important open issues concern the surface density of EROs and the
relative contribution of different classes of objects as a function of
both color and apparent magnitude. We continue to lack statistically large samples
of EROs. The most recent systematic large
survey, covering an area of 154~square arcminutes by Thompson et al. (1999),
has yielded six objects with $K \le 19.0$ and $ R - K > 6$.
To quantify the fraction of different classes of EROs as a function
of colors and magnitude, larger samples are required.
The ERO surface density inferred from the Thompson et al. survey is
0.04$\pm 0.016$ arcmin$^{-2}$ for $K \le 19.0$ and $R - K > 6$.
Depending on the K-band magnitude limit, $\rm R -
K$ color and possibly environment, the
reported surface densities of EROs range from $0.01 - 0.7$
arcmin$^{-2}$, derived from several serendipitous surveys over small
areas (Hu \&\ Ridgway 1994; Cowie et al. 1994; Beckwith et al.
1998; Thompson et al. 1999).
There have been suggestions that EROs tend to cluster, particularly,
in regions around high redshift AGNs compared with
blank fields.  The large
luminosities and (admittedly uncertain) space densities imply that these
objects represent a singificant consitutent of the overall galaxy
population and that their contribution to the overall rate of star
formation is non-negligable (e.g. Liu et al. 2000).

In this paper, we present a sample of EROs discovered using NICMOS on
HST while operating in the parallel mode. Our combined NICMOS/optical
survey covers only 16 square arcminutes but it provides high spatial
resolution and better signal-to-noise in the near-IR than most
ground-based surveys.

\medskip
\section{Observations and Reductions}
\bigskip

\subsection{Near Infrared Images}

The images under discussion here were obtained with camera 3 on NICMOS
(Thompson et al. 1998) using the F160W ($\lambda_c = 1.6\mu$m,
$\delta\lambda = 0.4\mu$m) and F110W band-passes ($\lambda_c =
1.1\mu$m, $\delta\lambda = 0.8\mu$m).  The data were obtained in the
parallel observing mode during the period from October 1997 to November
1998, excluding the dedicated NICMOS camera 3 observing campaigns.
This allowed us to observe fields which are essentially randomly
distributed over the sky. The sensitivity varies from field-to-field
due to different integration times and background levels.  The NICMOS
internal pupil adjustment mirror was set near the end of its travel,
providing the best possible focus for camera 3.  The PSF in the images
is slightly non-gaussian, but well characterized by FWHM$ = 0.25^{''}$. We
obtained four images per orbit, two each with the F110W
and F160W filters.  The field offset mirror (FOM) was used to
dither between two fixed
positions $1.8^{''}$ apart in a direction aligned with one axis of the
detector.  In addition there were small inter-orbit dither moves
executed for some of the pointings. The projected size of a camera 3
pixel is $0.204{''}$, giving a $51{''} \times 51^{''}$ field of
view for each image.  A small area is lost in the construction of the
final mosaic image.  The NICMOS camera 3 parallel program covered
approximately 200 square arcminutes at high galactic lattitude in its
14 months of operation.  In the present work, we are limited by the
amount and the depth of the visible-light data that we were able to
collect for the NICMOS parallel fields. As described below, most of the
NICMOS F160W images easily reach F160W magnitudes fainter than 21 (Vega magnitude)
with high signal-to-noise ratios (eg. 20$\sigma$ at H = 21.0).
Thus, to detect EROs at these depths  requires 3$\sigma$ R magnitude limits of
26 in our follow-up images.  As the NICMOS camera 3 field of view is
only 0.722 square arcminutes, optical followup is quite inefficient.

We used McLeod's (1997) NicRed v1.7 package to linearize and remove the
cosmic rays from the MultiAccum images.  Median images were derived
from more than 50 pointings and these were used to remove the dark and
sky signals.  Even with the optimal dark subtraction, there remain
considerable frame-to-frame variations in the quality of the final
images. The individual linearized, dark corrected, flatfielded and
cosmic ray cleaned images were shifted, masked and combined to produce
final mosaic images. Before shifting, each image was $2 \times 2$
block-replicated and integer ($0.1^{''}$) offsets were applied. In this
way we avoided smoothing or interpolation of the data.  The MultiAccum
process is not 100\% efficient in rejecting cosmic ray events and so we
applied a $3\sigma$ rejection when assembling the final mosaics.  More
details regarding the NICMOS imaging reduction can be found in Yan et
al. (1998).

In Table 1, we list $3\sigma$ surface brightness limits in the F160W filter for
the fields where we have obtained optical data.
The 50\%\ completeness depth in the
in the shallowest field is approximately m(F160W)$ = 22$.
Thus our near-IR catalog is 100\%\ complete
at the ERO selection limit (F160W $\sim 21.5$ as described below).
We emphasize that our ERO search is entirely limited by
the depth of the optical images.

\begin{deluxetable}{llrcccc}
\tablewidth{0pt}
\tablecaption{The Selected NICMOS Fields}
\tablehead{\colhead{Field} &
\colhead{RA} &
\colhead{DEC} &
\colhead{b} &
\colhead{T(F160W)} &
\colhead{$\mu(3\sigma)$} }
\tablecolumns{6}
\startdata
      & \multispan{2}\hfil (J2000) \hfil & (deg) & (sec) & mag/arcsec$^{2}$ & \\ \hline\hline
0041$+$3302 & 00:41:16 &  33:02:04 &  -30 &  2040  &  25.0  &  \\
0050$-$5200 & 00:50:12 & $-$52:00:01 &  -65 &  2040  &  25.1  &  \\
0049$-$5159 & 00:49:57 & $-$51:59:57 &  -65 &  2040  &  25.1  &    \\
0240$-$0140 & 02:40:01 & $-$01:40:39 & -54 & 2816 & 24.1 & \\
0240$-$0141a & 02:40:07 & $-$01:41:10 &  -54 &  2816  &  23.8 & \\
0240$-$0141b & 02:40:12 & $-$01:41:27 &  -54 &  2816  &  24.6 &  \\
0354$+$0943 & 03:54:51 &  09:45:09 &  -32 &  1020  & 24.2  &   \\
0354$+$0945 & 03:54:51 &  09:45:30 &  -32 &  1020  & 24.1 &   \\
0457$-$0456 & 04:57:19 & $-$04:56:51 &  -28 &  4480  &  25.7 &   \\
0729$+$6915 & 07:29:57 &  69:15:02 &   29 &  5120  &  25.9 &  \\
0741$+$6515 & 07:41:44 &  65:15:25 &   30 &  1030  &  24.9 &  \\
1504$+$0110 & 15:04:37 &  01:10:31 &   49 &   510  &  23.4  &    \\
1604$+$4318 & 16:04:55 &  43:18:56 &   48 &  3840  &  25.6 &     \\
1631$+$3001 & 16:31:39 &  30:01:23 &   42 &  3584  &  25.1 &    \\
1631$+$3730 & 16:31:40 & 37:30:30  & 43   & 3570   & 25.2 & \\
1631$+$3736 & 16:31:19 &  37:36:53 &   43 &  3570  & 24.9     &   \\
1657$+$3526 & 16:57:23 &  35:26:05 &   37 & 18360  &   25.4    &   \\
1940$-$6915 & 19:40:57 & $-$69:15:04 &   30 &  6120  &  23.9     &   \\
2044$-$3124 & 21:44:40 & $-$31:24:55 &   37 &  2040  &  24.9    &    \\
2220$-$2442 & 22:20:11 & $-$24:42:14 &   56 &  5110  &   25.7  &   \\
2325$-$2442 & 22:20:12 & $-$24:42:00 &  -56 &  5120  &  25.9 &    \\
2344$-$1524 & 23:44:00 & $-$15:24:50 &  -70 &  3840  &  25.4 &   \\
\enddata
\end{deluxetable}

\subsection{Optical Observations}

Multicolor CCD observations of several of the NICMOS parallel fields
were made at Cerro-Tololo, Las Campanas, WIYN, Palomar and the W. M.
Keck observatories.  The date of the observations, duration of the R
images, the FWHM of the seeing and the 3$\sigma$ limiting surface brightness
of the R images are listed in Table 2.  The pixel scales range from 
$0.2^{''}$ for the LCO, WIYN, and Keck cameras to $\sim 0.4^{''}$ for the CTIO
BTC camera system. Several of the fields were observed on more than
one occasions and from different sites to obtain photometric
calibration. The seeing was better than $1^{''}$ for most of the
fields. The data were reduced and calibrated using standard methods. The
resulting images were interpolated and rotated to match the NICMOS
fields and true color images were constructed from the F160W, F110W,
and R images (or F160W, R and V where possible). This allowed for easy
and efficient identification of objects with red colors. The photometry
was performed on the original uninterpolated images.

\begin{deluxetable}{ccccccc}
\tablewidth{0pt}
\tablecaption{Optical Follow-up Observations }
\tablehead{\colhead{Field} &
\colhead{Inst.} &
\colhead{Date} &
\colhead{FWHM} &
\colhead{T(R)} &
\colhead{$\mu(3\sigma)$}}
\tablecolumns{6}
\startdata
 &  & & arcsec & sec. &  mag/arcsec$^2$ & \\ \hline\hline
0041$+$3302 & WIYN & 11/25/98 & 0.9  & 3600  & 26.9  &  \\   
0050$-$5200 & LCO  & 10/14/98 & 1.1  & 3500  & 26.8  &   \\   
0240$-$0141 & CTIO & 09/15/98 & 1.2  & 2500  & 26.8  &   \\   
0354$+$0945 & LCO  & 10/12/98 & 0.8  & 4000  & 26.6  &    \\
0457$-$0456 & LCO  & 03/02/98 & 0.8  & 3000  & 26.2  &    \\
0729$+$6915 & WIYN & 11/25/98 & 0.8  & 4800  & 26.9  & \\
0741$+$6515 & WIYN & 01/18/99 & 0.7  & 3600  & 27.1  &  \\
1504$+$0110 & WIYN & 05/03/98 & 0.8  & 1800  &  25.8  &   \\
1604$+$4318 & Keck & 06/22/98 & 0.6  & 1800  & 26.6   & cirrus   \\
1604$+$4318 & WIYN & 05/02/98 & 0.7  & 3000  & 26.5   &    \\
1631$+$3001 & Keck & 05/21/99 & 0.7  &  900  &  27.3 &   \\
1631$+$3001 & WIYN & 04/27/98 & 0.7  & 3600  & 26.2  & \\
1631$+$3736 & Keck & 05/21/99 & 0.6  &  300  & 26.5  &  \\
1657$+$3526 & Keck & 05/21/99 & 0.7  &  600  &  27.0 &  \\
1940$-$6915 & CTIO & 09/14/98 & 1.1  & 2500  &  26.8 &  \\
2044$-$3124 & LCO  & 10/13/98 & 0.9  & 3500  &  26.7 &    \\
2220$-$2442 & LCO  & 10/12/98 & 0.8  & 5400  &  27.1 &   \\
2325$-$2442 & WIYN & 10/24/98 & 0.8  & 1800  &  26.1 &   \\
2344$-$1524 & CTIO & 09/13/98 & 1.0  & 4000  &  26.9 &  \\
\enddata
\end{deluxetable}

\begin{figure*}
\plotone{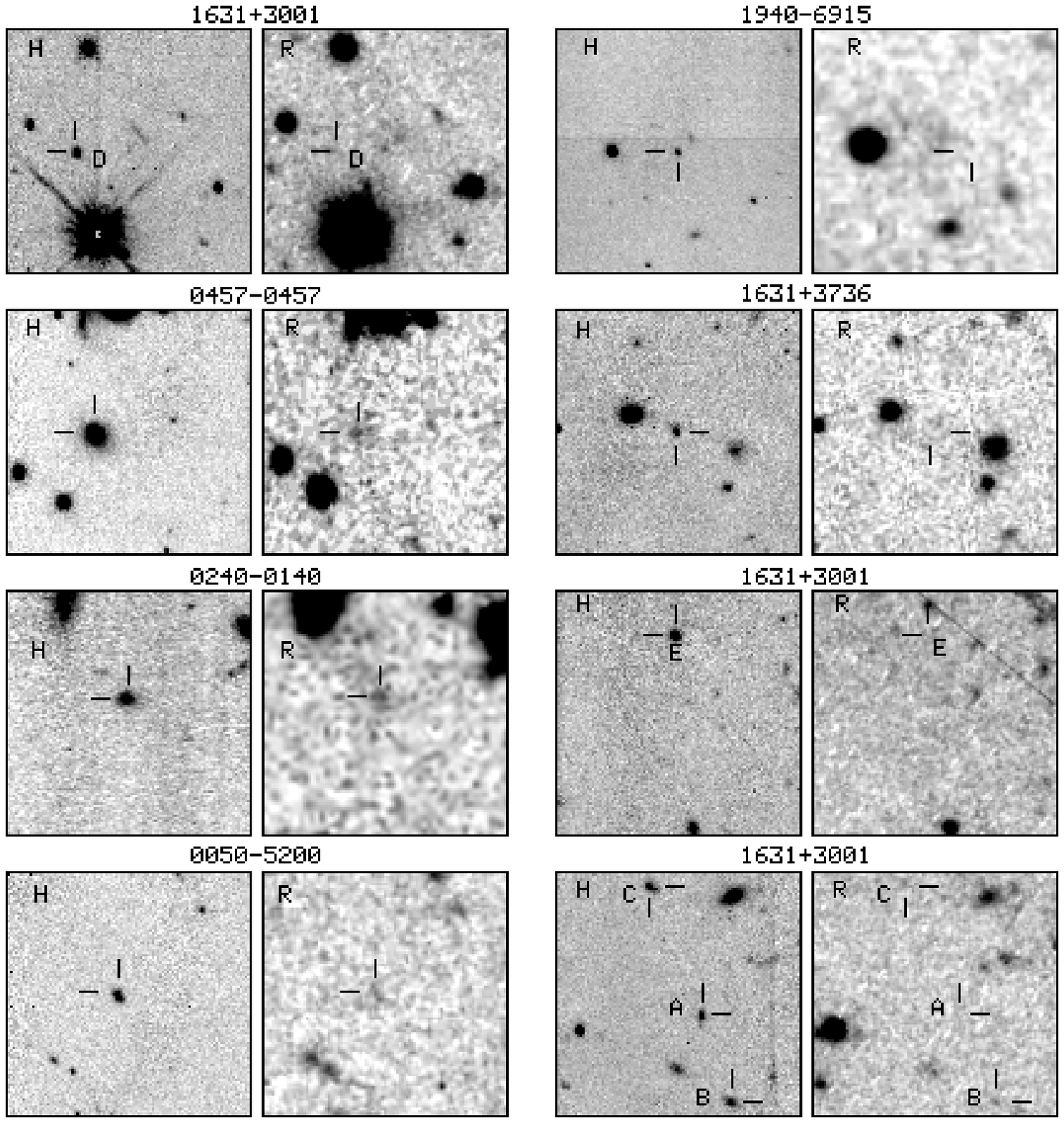}
\end{figure*}

\begin{figure*}
\plotone{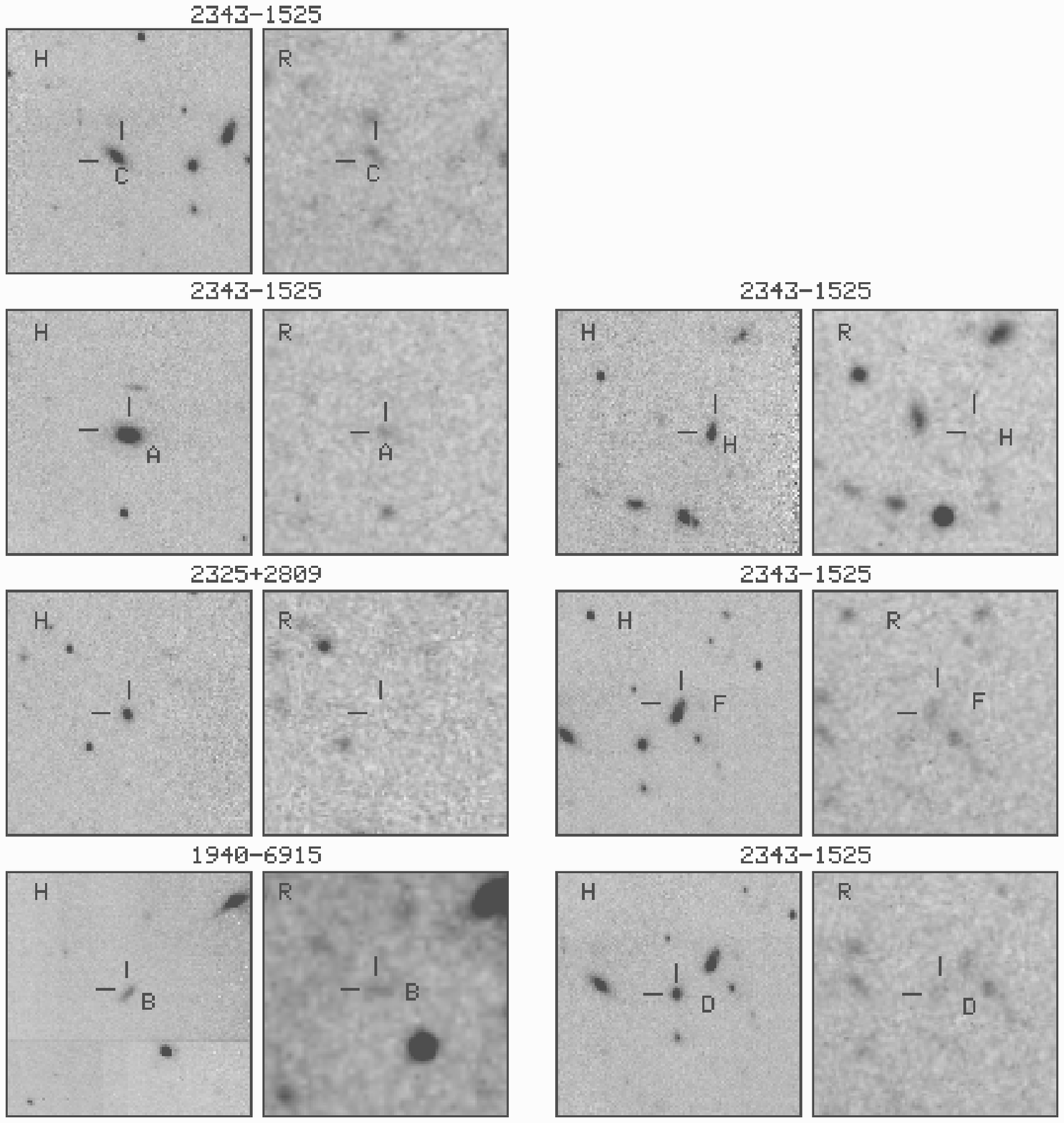}
\caption{Images of EROs in the R and H bands.}
\end{figure*}

\section{Results}
\bigskip

\subsection{ERO Detection and Photometry}

All of the ERO identifications were made visually from comparisons
between the aligned near-IR and optical images. The small size of each
NICMOS image ($51''\times 51''$) as well as its sensitive depth made
visual selection effecient.  All of the EROs are well detected in the
near-IR bands and were either undetected or marginally detected in the
R-band images. ($\rm R - F160W$) colors were measured for all of the 
ERO candidates using the original uninterpolated images. The colors are
subject to significant uncertainties due to the low signal levels of
the red objects in the R-band images. The mismatch between the point
spread functions of NICMOS and ground-based CCD images adds a
significant complication to the color measurement.
We adopted the following method to estimate the ($\rm R - F160W$) colors.
We smoothed our F160W images to the same FWHM as the corresponding
optical images. The F160W and R magnitudes were then measured in
identical apertures. The aperture diameter is
$2.5\times$ the measured FWHM in the R-band images.  We used the Source
Extractor software (Bertin \&\ Arnouts 1996) to measure the
magnitudes.

In Table 3 we also list the magnitudes of EROs in a variety of filters.
These are isophotal magnitudes measured out to 1$\sigma$ isophotal
radius.  As our sample of EROs are well detected at F160W, isophotal
magnitudes are a good measure of total magnitudes.
For the EROs which are undetected in the optical 
images, their 3$\sigma$ magnitude limits within a 2.5FWHM
diameter aperture are used.

\begin{deluxetable}{lccrrrrrrrrrc}
\tablecolumns{12}
\tablewidth{0pt}
\tablecaption{The EROs in the NICMOS Parallel Fields}
\tablehead {
\colhead{Obj. Name}  &
\colhead{RA$_{J2000}$ } &
\colhead{DEC$_{J2000}$} &
\colhead{R$-$K}   &
\colhead{R$-$H} &
\colhead{K}   &
\colhead{H}   &
\colhead{J}  &
\colhead{I}   &
\colhead{R}   &
\colhead{V}   &
\colhead{B}  }
\startdata
0050$-$5200A & 00:50:13.176 & $-$52:00:12.62 & ... & $>$5.2 & ... & 20.6 & 22.4 & $>$23.8& $>$25.9 & $>$25.6& ... & \\
0240$-$0140A & 02:40:06.780 & $-$01:41:40.76 & ... & 5.1 & ... & 19.2 & 20.9 & ... & 24.3 & $>$26.5 & $>$26.3& \\
0457$-$0457A & 04:57:20.184 & $-$04:57:12.46 & 6.6 & 6.0 & 18.17 & 18.8 & 20.5 & 23.2 & 24.8 & $>$25.0 & \\
1631$+$3001A & 16:31:39.180 & +30:01:19.60 & ... & $>$5.3 & ... & 21.4 & 23.1 & ... & $>$26.9 & ...& ... &\\
1631$+$3001B & 16:31:38.885 & +30:01:26.33 & ... & $>$5.6 & ... & 21.3 & 23.2 & ... & $>$26.9 & ...& ... & \\
1631$+$3001C & 16:31:39.609 & +30:01:09.42 & ... & $>$5.5 & ... & 21.2 & 22.7 & ... & $>$26.9 & ...& ... & \\
1631$+$3001D & 16:31:38.197 & +30:01:01.51 & ... & $>$6.1 & ... & 20.8 & 22.3 & ... & $>$26.9 & ...& ... & \\
1631$+$3001E & 16:31:40.741 & +30:00:49.19 & ... & $>$6.3 & ... & 20.3 & 22.0 & ... & $>$26.9 & ...& ... & \\
1631$+$3736A & 16:31:19.549 & +37:37:01.08 & ... & 5.0 & ... & 21.4 & 23.6 & ... & 25.0 & ... & ... & \\
1940$-$6915A & 19:40:55.732 & $-$69:15:17.82 & ... & $>$5.2 & ... & 20.5 & 22.1 & ... & $>$25.7 & $>$25.1 & $>$26.0 & \\
1940$-$6915B & 19:40:57.821 & $-$69:15:07.40 & ...& $>$5.2 & ... & 20.6 & 22.6 & ... & $>$25.7 & $>$ 25.1 & $>$26.0 & \\
2325$+$2809A & 23:25:01.830 & +28:09:25.30 & ... & $>$5.3 &... & 20.2 & 22.2 & ...& $>$25.5 &... & ...& \\
2344$-$1525A & 23:43:58.504 & $-$15:25:12.74 & 6.1 & 5.0 & 17.8 & 18.9 & 20.8 & ... & 23.9 & 25.6 & $>$26.2& \\
2344$-$1525C\tablenotemark{a} & 23:43:58.675 & $-$15:24:56.33 & 5.6 & 4.4 & 18.7 & 19.9 & 21.5 & ... & 24.3 & 24.6 & $>$26.2& \\
2344$-$1525D & 23:43:59.103 & $-$15:24:57.42 & $>$5.8 & $>$5.0 & 19.5 & 20.5 & 21.8 & ... & $>$25.4 & $>$25.6 & $>$26.2 & \\
2344$-$1525F\tablenotemark{a} & 23:43:59.254 & $-$15:25:00.63 & 5.9 & 4.5 & 18.2 & 19.7 & 21.4 & ... & 24.1 & 24.2 & 24.9 & \\
2344$-$1525H & 23:44:01.113 & $-$15:25:10.83 & $>$6.2& $>$5.2 & 19.1 & 20.2 & 21.9 & ... & $>$25.4& $>$25.6 & $>$26.2 & \\
\enddata
\tablenotetext{a}{For these two galaxies their R $-$ H colors are slightly less
than 5, but since they belong to the same cluster group, we included them for the completeness.
These two objects are excluded in the ERO surface density calculation.}
\end{deluxetable}

\subsection{Surface Density}
In a total of 22 NICMOS fields, 
covering 16 square arcminutes,
we detect 15 objects with $\rm R - F160W > 5$.  Table 3 lists
their equatorial coordinates, near-IR and optical magnitudes, and
optical-IR colors. For some fields, we also have K-band images taken
with NIRC (Matthews \&\ Soifer, 1994) 
on the Keck telescope (Teplitz et al. 2000). 
The ERO sample is selected  with $\rm R - F160W >5$
without any limits set on H magnitude. 
Our color selection system is similar to $\rm R - H$ used 
in ground-based obsevations, and so direct comparisons are possible.

Figure 1 shows the R and F160W band images of each ERO.  
All of our EROs are resolved in the NIC3 images and appear to be extended,
except ERO 1940-6915A. This object has m(F160W) =  20.6.
and appears to be unresolved in our NIC3 image.
The reamining objects exhibit both elliptical and indeterminate morphologies.
The detailed analyses of the ERO luminosity profiles using both 
the NICMOS camera 2 and camera 3 data will be presented in Yan \&\ McCarthy (2000).

In a few fields where we have very
deep optical images taken with LRIS (Oke et al. 1995)
at the Keck 10~m telescope, we were able to
find EROs with H magnitude as faint as 21.4. The surface density of
EROs depend on the magnitude limit of the sample.  
Including the two apparent clusters of EROs and the point source like object ERO~1940-6915A, 
we estimate surface
densities of 0.19$\pm 0.11$~arcmin.$^{-2}$, 0.19$\pm 0.11$~arcmin.$^{-2}$,
0.50$\pm 0.18$~arcmin.$^{-2}$, and 0.94$\pm 0.24$ for EROs with $\rm R - F160W
>5$ and $\rm F160W < 19.5$, $\rm F160W < 20.0$,  $\rm F160W < 20.5$, and
$< 21.5$ respectively.  
If we exclude the two ERO clusters and the possible stellar object, the
inferred surface density is significantly lower, 0.13$\pm 0.08$~arcmin.$^{-2}$, 
0.13$\pm 0.08$~arcmin.$^{-2}$, 0.25$\pm 0.13$~arcmin.$^{-2}$ and 
0.38$\pm 0.15$~arcmin.$^{-2}$
for EROs brighter than 19.5, 20, 20.5 and 21.5
respectively. Here we did not include 2344$-$1525C and 2344$-$1525F whose R $-$ H
colors are slightly less than 5. Since our areal 
coverage is very small, the statistical errors in these
estimates are large, particularly at bright magnitudes. 
If EROs have F160W - K colors of $0.5 - 1$ as suggested by our observations,
our surface density estimate is consistent within 3$\sigma$ 
with that inferred from the ground-based ERO
survey by Thompson et al. (1999) of 0.04$\pm 0.016$~arcmin.$^{-2}$ for R $-$ K $>$ 6 and 
K $<$ 19.




\subsection{Infrared Colors}
In Figure 2, we present F110W - F160W colors for all of the galaxies
detected in 19 NICMOS fields, where we have both F160W and F110W images
as well as ground-based optical follow-up observations (see Table 1).
The solid dots in the figure represent
the EROs listed in Table 3. Overlaying on top of the data are 
color-magnitude tracks for a passively evolving L$^\star$ elliptical galaxy
formed at $z=10$ in a single burst lasting 1~Gyr and 
a star-forming galaxy with a constant star formation rate 
of 1~M$_\odot$/yr respectively. Here we adopt H$\rm _0=70~kms^{-1}Mpc^{-1}$
and q$_0=0.1$. The open diamonds indicate the places with 
redshifts of 0.5, 1, 2, and 3. The error bars on F110W - F160W colors
and F160W magnitudes are $\pm 1\sigma$, calculated from Sextractor (Bertin \&\ Arnouts 1996).
The F110W $-$ F160W colors
of our R - F160W selected EROs appear to be redder than the average color of field galaxies
(Figure 2). They also appear to be redder than the $z \sim 0.9$ cluster elliptical
population studied by Stanford et al. (1998) (J $-$ H $\sim 1.1$). 
There are some objects which have red J $-$ H colors, but 
R $-$ H colors that do not meet our ERO definition. These objects may contain
small amounts of current star formation.  If the objects in our ERO sample have
luminosities near L$^*$, the median redshift of our sample is $\sim 1$, and the
total range sampled is roughly $ 0.6 < z < 1.5$.  

Among the 22 NICMOS pointings for which we have deep optical photometry,
we found two fields that each contains more than 3 objects meeting our ERO definition. 
Plate 1 and 2 show the VR~F160W composite true color pictures of the ERO groups in 
2344$-$1524 and 1631+3001. In the 2344$-$1524 field, we also obtained K band 
images and the selected ERO candidates have $\rm R - K \gta 6$.
The 1631$+$3001 field has many red objects, including several very faint objects with 
$\rm F(160W) > 21.5$, which are not selected in our ERO sample.
Plates 1 and 2 clearly indicate two potential groups or clusters of EROs. 
Follow-up spectroscopy would
be important for determining the cluster redshifts and the physical nature of these
EROs. Our detections of two clusters of EROs over a small area of $\sim 16$ sqaure arcminutes
suggest that EROs tend to be strongly clustered. Previous studies by 
McCarthy et al. (1992) and Thompson et al (1999) have also noted the ERO clustering 
phenomenon from small statistical samples. Obviously, the problem of 
ERO clustering will be an important goal for future deep infrared surveys covering large
area.

\section{ Acknowledgments}
We thank the staff of the Space Telescope Science Institute for their
efforts in making this parallel program possible. In particular we
thank Duccio Machetto, Peg Stanley, Doug van Orsow, and the staff of the PRESTO
division. We also thank John Mackenty and members of the STScI  NICMOS
group for crafting the exposure sequences. Some of the data
presented herein were obtained at the W.M. Keck Observatory,
which is operated as a scientific partnership among the
California Institute of Technology, the University of California
and the National Aeronautics and Space Administration.  The
Observatory was made possible by the generous financial
support of the W.M. Keck Foundation.  This research
was supported, in part, by grants from the Space Telescope Science
Institute, GO-7499.01-96A, AR-07972.01-96A and PO423101.  HIT acknowledges funding
by the Space Telescope Imaging Spectrograph Instrument Definition Team
through the National Optical Astronomy Observatories and by the NASA
Goddard Space Flight Center.


\noindent{Plate 1. --- This is a VR~F160W composite
color image of 2343$-$1524 field. The field shown here has a size of 
50$^{''}\times$50$^{''}$.  The color image indicates a cluster of EROs.}

\noindent{Plate 2. --- A VR~F160W composite
color picture of 1631+3001 field shows a cluster of EROs.
The field size is the same as in Plate 1.}

\begin{figure*}
\plotone{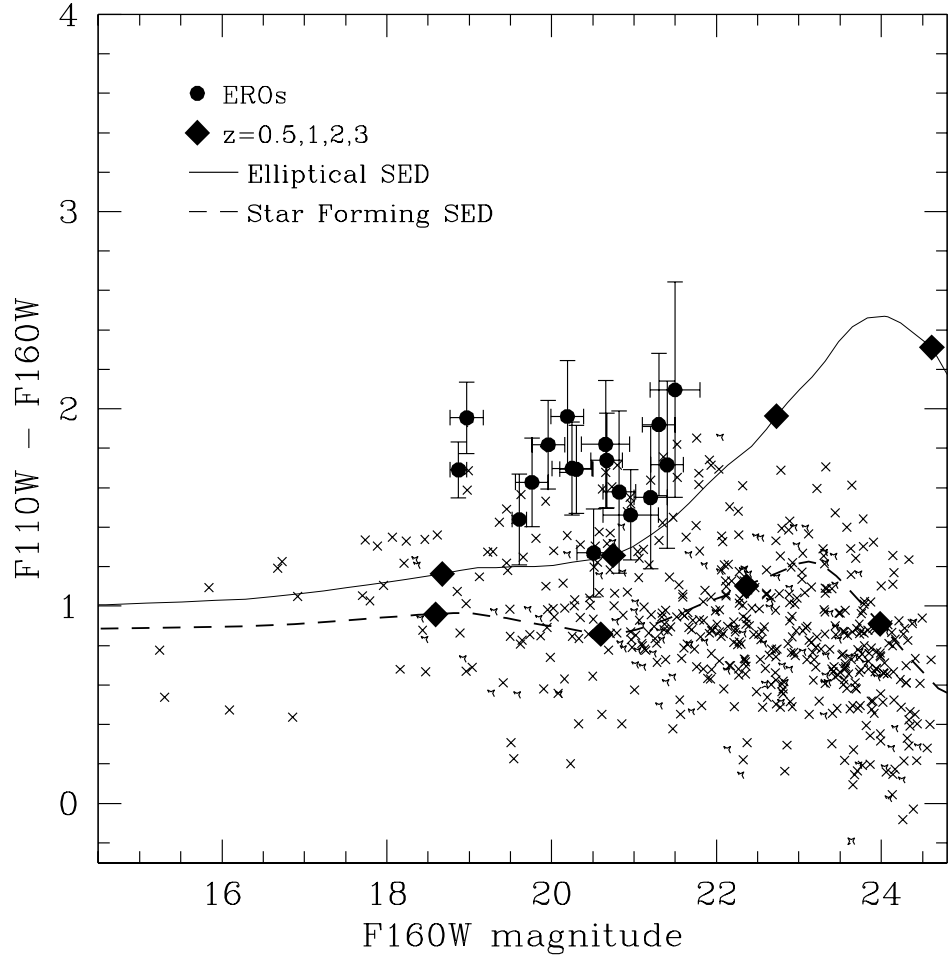}
\caption{Infrared Color-magnitude diagram for galaxies detected
in 19 NICMOS fields. The solid dots show the EROs listed in Table 3.
The solid and dashed lines are synthetic color-magnitude tracks
for an old elliptical and a star-forming galaxy using Bruzual
\&\ Charlot model. The open diamonds indicate the places where redshifts
are equal to 0.5, 1., 2., and 3. Here we assume H$_0=70~kms^{-1}Mpc^{-1}$ and q$_0=0.1$.
The errorbars are $\pm 1\sigma$.}
\end{figure*}

\end{document}